\documentclass[%
 twocolumn,
 11pt,
 a4,
 reprint,
 superscriptaddress,
 notitlepage,
 amsmath,amssymb,
 aps,
 longbibliography,
 floatfix,
]{revtex4-2}

\usepackage[table]{xcolor}
\usepackage{graphicx}
\usepackage{dcolumn}
\usepackage{bm}
\usepackage{physics}
\usepackage{appendix}
\usepackage{float}
\usepackage{hyperref}
\usepackage[percent]{overpic}
\usepackage{subcaption}
\usepackage[font=small,labelfont=bf]{caption}
\begin{document}

\title{Generative Circuit Design for Quantum-Selected Configuration Interaction}
\date{\today}

\begin{abstract}
Quantum-selected configuration interaction (QSCI) has emerged as a feasible approach for approximating electronic ground states on noisy quantum devices toward large-system demonstrations. In QSCI, Slater determinants are sampled from a quantum-prepared state, and the Hamiltonian is then diagonalized in the sampled subspace. To create a high-quality subspace under hardware constraints, the design of the state-preparation circuit is crucial. Here, we present a Generative Quantum Eigensolver (GQE)-based framework that optimizes ansatz structures using a Transformer policy trained on the QSCI subspace energy. We validate the framework on $\mathrm{N}_2$ in active spaces of up to 32 qubits. We found that the optimized circuits reach chemical precision with substantially lower gate counts than time-evolved circuits. Quantitatively, this corresponds to an average reduction of 98\% in the required two-qubit gate count relative to the single-step first-order Trotterized approximation and 83\% relative to the qDRIFT approximation. Furthermore, the resulting wavefunctions are competitive with heat-bath configuration interaction (HCI) in terms of compactness. In stretched-bond, strongly correlated regimes, they achieve chemical precision with subspaces that are 50\% smaller than those required by HCI.

\end{abstract}

\author{Ryota Kemmoku}
\thanks{ryota.kemmoku@keio.jp}
\affiliation{Department of Applied Physics and Physico-Informatics, Keio University, Hiyoshi 3-14-1,
Kohoku, Yokohama 223-8522, Japan}

\author{Qi Gao}
\affiliation{Mitsubishi Chemical Corporation, Science \& Innovation Center,
1000, Kamoshida-cho, Aoba-ku, Yokohama 227-8502, Japan}
\affiliation{Quantum Computing Center, Keio University, Hiyoshi 3-14-1,
Kohoku, Yokohama 223-8522, Japan}

\author{Shu Kanno}
\affiliation{Mitsubishi Chemical Corporation, Science \& Innovation Center,
1000, Kamoshida-cho, Aoba-ku, Yokohama 227-8502, Japan}
\affiliation{Quantum Computing Center, Keio University, Hiyoshi 3-14-1,
Kohoku, Yokohama 223-8522, Japan}

\author{Kimberlee Keithley}
\affiliation{Mitsubishi Chemical Corporation, Science \& Innovation Center,
1000, Kamoshida-cho, Aoba-ku, Yokohama 227-8502, Japan}
\affiliation{Quantum Computing Center, Keio University, Hiyoshi 3-14-1,
Kohoku, Yokohama 223-8522, Japan}

\author{Ikko Hamamura}
\affiliation{
NVIDIA, ATT EAST 12F, 2-11-7 Akasaka, Minato-ku, Tokyo 107-0052, Japan
}

\author{Naoki Yamamoto}
\affiliation{Department of Applied Physics and Physico-Informatics, Keio University, Hiyoshi 3-14-1,
Kohoku, Yokohama 223-8522, Japan}
\affiliation{Quantum Computing Center, Keio University, Hiyoshi 3-14-1,
Kohoku, Yokohama 223-8522, Japan}

\author{Kouhei Nakaji}
\affiliation{
NVIDIA, Santa Clara, California, USA
}

\maketitle

\section{Introduction}
Quantum computing holds promise for broadening the range of tractable problems in quantum chemistry. In the long term, algorithms based on quantum phase estimation~\cite{Kitaev1995QPE,AbramsLloyd1999QPE,AspuruGuzik2005QPE} offer a direct way to obtain eigenvalues of molecular Hamiltonians. However, their circuit depth and coherence requirements remain far beyond what current and near-term hardware can support. This gap has led to strong interest in hybrid quantum--classical algorithms for noisy or early-fault-tolerant devices~\cite{katabarwa2024early,preskill2025beyond}.

In this context, quantum-selected configuration interaction (QSCI) ~\cite{Kanno2023QSCI}, also referred to as sample-based quantum diagonalization (SQD)~\cite{RobledoMoreno2025ScienceAdv}, has emerged as a particularly attractive framework. The central idea is to use a quantum device not to estimate the full expectation value of the Hamiltonian, but to prepare an input state and sample computational-basis bitstrings from it. In quantum chemistry, these bitstrings correspond to Slater determinants, which define a reduced subspace in which the electronic Hamiltonian is then diagonalized classically. In this sense, QSCI can be viewed as a quantum extension of classical selected-CI approaches such as the CI using a perturbative selection made iteratively (CIPSI) method ~\cite{Huron1973CIPSI, Evangelisti1983CIPSI} and heat-bath configuration interaction (HCI)~\cite{Holmes2016HCI,Sharma2017SHCI}. The distinctive promise of QSCI lies in its use of quantum sampling to access determinant distributions that may be classically hard to generate. This promise is realized only when the prepared quantum state induces a distribution that efficiently exposes significant configurations.

How to design such an input state remains unresolved. A widely used choice is the local unitary cluster Jastrow (LUCJ) ansatz~\cite{Matsuzawa2020Jastrow,Motta2023LUCJ}, which embeds a coupled cluster singles and doubles (CCSD) ~\cite{Bartlett2007CCReview} wavefunction into a local approximation that balances chemical motivation with hardware friendliness. Its practicality has been demonstrated in large-scale SQD studies, including a 77-qubit calculation for the [4Fe-4S] cluster~\cite{RobledoMoreno2025ScienceAdv}, and the same framework has since been extended to intermolecular interactions~\cite{Kaliakin2025Intermolecular}, excited-state calculations~\cite{Barison2025ExtendedSQD,Liepuoniute2025Methylene}, and combinations with density matrix embedding theory~\cite{Shajan2025DMETSQD}. However, it remains unclear whether an LUCJ-prepared state provides a sufficiently good approximation to the true ground state for QSCI. In particular, recent work has pointed out that QSCI based on the LUCJ ansatz can suffer from poor sample efficiency, due to repeated measurements failing to uncover new significant determinants, and from noncompact wavefunctions compared to classical selected-CI references~\cite{Reinholdt2025FatalFlaw}.

An alternative major route is to generate the QSCI input state through Hamiltonian time evolution. This approach was initially proposed by Sugisaki et al.~\cite{Sugisaki2025HSBQSCI} as Hamiltonian simulation-based QSCI, and later employed by Mikkelsen and Nakagawa~\cite{Mikkelsen2025TEQSCI}, Yu et al.~\cite{Yu2025SKQDv3}, the latter of whom formalized it within the framework of a quantum Krylov subspace method~\cite{Cortes2022QKSD,Zhang2024QKSD}. The motivation behind these approaches is that, if the initial state has sufficient overlap with the true ground state, then time-evolved states can expose the configurations needed to build an effective Krylov-inspired subspace, giving a principled route to systematic improvement. However, the practical difficulty is circuit depth. Since the number of Hamiltonian terms grows polynomially with system size in electronic-structure problems, even a very short Trotterized evolution can already become too costly for present hardware. The recently proposed SqDRIFT~\cite{Piccinelli2025SqDRIFT,campbell2018random,wan2022randomized} framework significantly reduces this cost by randomizing the propagator. However, recent numerical studies have shown that the resulting wavefunctions can still be less compact than HCI references, as observed in 48-qubit calculations for coronene~\cite{Piccinelli2025SqDRIFT} and 42-qubit calculations for SiH$_4$~\cite{Weaving2025Compact}.

These limitations have motivated a different strategy: instead of fixing the input-state family in advance, one attempts to optimize it. The most straightforward realization of this idea is by using the variational quantum eigensolver (VQE)~\cite{Peruzzo2014VQE,McClean2016TheoryVQE,Bittel2023VQAHardness,Cai2020ObservableFrugal,Franca2021LimitsVQA,Cerezo2021BP} as in the original QSCI proposal~\cite{Kanno2023QSCI} and in recent hardware demonstrations~\cite{Nutzel2025IndustrialVQE}. Such studies demonstrate that improving state preparation can enhance the quality of sampled subspaces. At the same time, standard variational optimization often suffers from barren plateaus in expressive parameter landscapes.

The generative quantum eigensolver (GQE)~\cite{Nakaji2025GQE} is a recently proposed framework to address this difficulty. By replacing optimization over continuous circuit parameters with a combinatorial search over a discrete operator pool, GQE is expected to avoid barren-plateau problem. Concretely, a Transformer-based policy~\cite{Vaswani2017Transformer} generates candidate operator sequences, whose performance is then used to update the policy toward more effective circuit structures~\cite{Nakaji2025GQE}. Other approaches to circuit optimization have also been explored, including ADAPT-QSCI~\cite{Nakagawa2024ADAPTQSCI}, HI-VQE~\cite{PellowJarman2025HIVQE,Yoo2026ExtendHIVQE}, and closed-loop optimization based on differential evolution~\cite{Shirakawa2025DE}. However, GQE offers two features that are particularly attractive for QSCI. First, because it optimizes the circuit structure itself rather than only the circuit parameters, it can in principle optimize the circuit cost together with the quality of the resulting sampled subspace. Second, because the search policy is learned, it opens the possibility of reusing search bias across related molecular systems. If molecule-to-molecule or size-to-size transfer becomes feasible, then ansatz optimization need not be restarted from scratch for every new problem, and a substantial reduction in optimization cost may become possible.

In this work, we adapt the GQE framework to QSCI and use it to optimize the ansatz structure under explicit constraints on both circuit depth and the maximum dimension of the classical diagonalization subspace, $d_{\text{max}}$. We further introduce a refinement pipeline, inspired by carryover ideas explored in HI-VQE \cite{PellowJarman2025HIVQE,Yoo2026ExtendHIVQE} and closed-loop optimization workflows \cite{Shirakawa2025DE}, in which approximate wavefunctions generated during the optimization process are used to identify and accumulate significant determinants across iterations. This allows the method to combine learned ansatz search with progressive subspace refinement. In numerical experiments on N$_2$, we compare sampling efficiency, gate cost, and wavefunction compactness in the 16-qubit active space and additionally examine scaling up to 32 qubits. We find that the optimized circuits reach chemical precision with substantially lower gate counts than time-evolved baselines, reducing the required two-qubit gate count by an average of 98\% relative to the single-step first-order Trotterized approximation and by 83\% relative to the qDRIFT approximation. We further find that, at large N--N separations where strong static correlation dominates, the refinement procedure yields subspaces that are not only more accurate but also more compact than HCI, reaching chemical precision with roughly 50\% fewer determinants.

The remainder of this paper is organized as follows. Section~\ref{sec:methods} introduces the proposed framework, beginning with a brief review of QSCI and then describing the GQE-based ansatz optimization and the local and the global refinement procedures. Section~\ref{sec:result} presents numerical results, including the optimization performance, comparisons of sampling efficiency, gate efficiency, and wavefunction compactness with existing QSCI ansatz families, and system-size scaling at fixed shot count. Section~\ref{conclusion} concludes with a discussion of the main findings and future directions.

\begin{figure*}[!t]
    \centering
    \includegraphics[width=\textwidth,page=1]{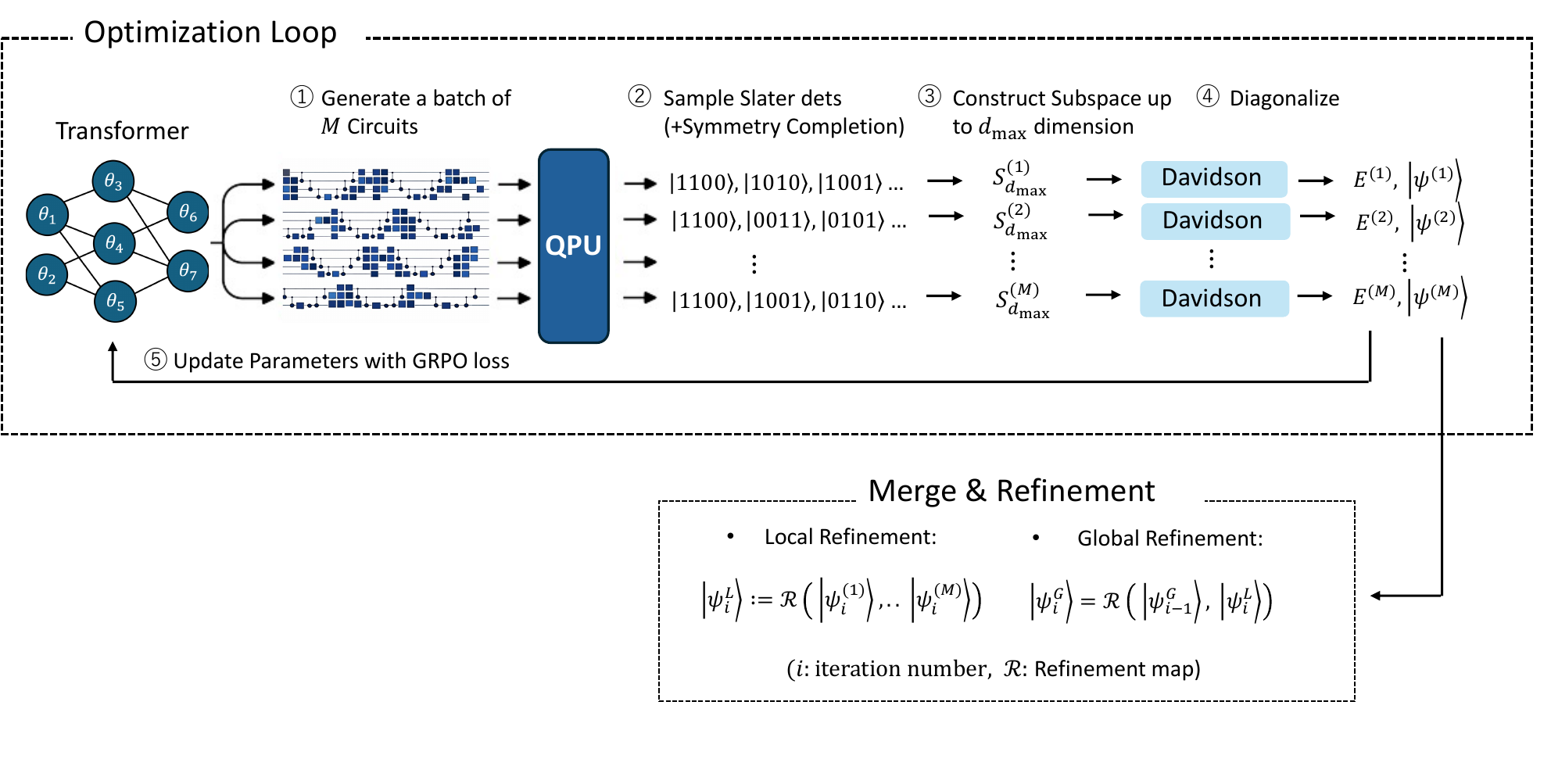}
    \caption{\textbf{Overall workflow of the proposed method.} At each optimization iteration, a Transformer policy generates $M$ candidate circuits. Each circuit is executed on the quantum device, measured in the computational basis, symmetry-completed, truncated to at most $d_{\max}$ determinants, and then classically diagonalized. The resulting subspace energies $E^{(m)}$ are used as rewards for the policy update, while the approximate QSCI wavefunctions $\ket{\psi^{(m)}}$ are passed to the refinement workflow described in Sec.~\ref{sec:refinement}.}
    \label{fig:workflow}
\end{figure*}

\section{Methods}\label{sec:methods}
\subsection{Quantum-selected configuration interaction}\label{subsec:qsci}
Quantum-selected configuration interaction (QSCI)~\cite{Kanno2023QSCI,RobledoMoreno2025ScienceAdv} is a quantum--classical hybrid method in which a quantum device is used to identify a determinant subspace and the Hamiltonian is classically diagonalized in that subspace. Under the Jordan--Wigner encoding, a computational-basis bitstring is in one-to-one correspondence with a Slater determinant. We denote such a determinant by
\begin{equation}
    \ket{x}=\bigotimes_{p=1}^{N_{\mathrm{orb}}}\ket{x_p^\uparrow}\ket{x_p^\downarrow},
    \qquad
    x_p^\sigma\in\{0,1\},
\end{equation}
where $p$ labels spatial orbitals and $\sigma\in\{\uparrow,\downarrow\}$ is the spin. We also write $x=x^\downarrow\oplus x^\uparrow$ when the two spin sectors need to be separated.

Let $\ket{\Psi}=\sum_x a_x\ket{x}$ be the state prepared on the quantum device. Measuring $\ket{\Psi}$ in the computational basis yields determinants distributed as $p(x)=|a_x|^2$. In practice, we perform $N_{\mathrm{shot}}$ measurements and count the number of occurrences of each sampled determinant $x$. From the resulting histogram, we form a candidate set $\widetilde{\mathcal D}$ and select up to $d_{\max}$ determinants in descending order of sampling frequency. These $d_{\max}$ determinants form
\begin{equation}
    \mathcal D_{d_{\max}}\subseteq \widetilde{\mathcal D},
    \qquad
    |\mathcal D_{d_{\max}}|\le d_{\max},
\end{equation}
and the associated QSCI subspace
\begin{equation}
    S_{d_{\max}}:=\mathrm{span}\{\ket{x}\mid x\in \mathcal D_{d_{\max}}\}.
\end{equation}
Restricting the Hamiltonian to this sampled subspace, we obtain the projected Hamiltonian
\begin{equation}
    \hat H_{d_{\max}}:=\hat P_{d_{\max}}\hat H\hat P_{d_{\max}},
    \qquad
    \hat P_{d_{\max}}:=\sum_{x\in \mathcal D_{d_{\max}}}\ket{x}\bra{x}.
\end{equation}
The energy is then obtained by solving the lowest-eigenvalue problem of $\hat H_{d_{\max}}$ classically, typically using the Lanczos or Davidson method:
\begin{equation}
    \hat H_{d_{\max}}\ket{\psi}=E\ket{\psi},
    \qquad
    \ket{\psi}=\sum_{x\in \mathcal D_{d_{\max}}} a_x\ket{x},
\end{equation}
where $E$ is the lowest eigenvalue. Because this is the Rayleigh--Ritz minimum within the sampled subspace, the energy remains variational.

\subsection{Generative quantum eigensolver for QSCI}
The Generative Quantum Eigensolver (GQE)~\cite{Nakaji2025GQE} is a reinforcement learning-based framework for ground-state search that treats ansatz construction as a generative modeling problem. Unlike the variational quantum eigensolver (VQE), which optimizes continuous parameters within a predetermined circuit family, GQE learns a probability distribution over discrete operator sequences drawn from a predefined operator pool. A generative model then samples candidate circuits, executes them on a quantum device, and updates the model parameters from energy-based rewards so that higher-quality circuit structures are sampled more frequently.

In this work, we combine GQE with QSCI by using the sampled circuit not as the variational ansatz itself, but as a generator of determinants for QSCI. More specifically, we replace the Hamiltonian expectation value used in the original GQE with the energy obtained after classical diagonalization in the sampled subspace. The overall workflow is summarized in Figure~\ref{fig:workflow}. Below, we describe circuit generation and policy optimization in this order.

\subsubsection{Transformer-based circuit generation}
Let $\mathcal G=\{U_j\}_{j=1}^{|\mathcal G|}$ be the operator pool, whose detailed construction is deferred to Appendix~\ref{app:operator_pool}. We represent the circuit-generation policy by a decoder-only Transformer~\cite{Vaswani2017Transformer} with parameters $\theta$, and denote the resulting autoregressive policy by $\pi_\theta$. The policy samples an operator-index sequence
\begin{equation}
    s=(s_1,\ldots,s_L),
    \qquad
    s_t\in\{1,\ldots,|\mathcal G|\},
\end{equation}
according to
\begin{equation}
    \pi_\theta(s\mid q)
    =
    \prod_{t=1}^{L}
    \pi_\theta(s_t\mid q,s_{<t}),
\end{equation}
where $t \in \{1,\dots, L\}$, $q$ is the start token, $L$ is the circuit length, and $s_{<t}:=(s_1,\dots,s_{t-1})$ denotes the previously sampled operator indices. Each sequence defines the circuit
\begin{equation}
    U(s):=U_{s_L}\cdots U_{s_1},
\end{equation}
and the corresponding sampled state
\begin{equation}
    \ket{\Psi(s)}:=U(s)\ket{\Phi},
\end{equation}
where $\ket{\Phi}$ is an initial state, and we adopt the Hartree--Fock determinant $\ket{\Phi_{\mathrm{HF}}}$ as $\ket{\Phi}$. 

\subsubsection{Reward evaluation and policy update}
Starting from a randomly initialized policy, we repeat the following procedure for $N_{\mathrm{iter}}$ iterations. At each iteration, the Transformer samples a batch of $M$ circuits, $\{U(s^{(m)})\}_{m=1}^{M}$, from the current policy. Here and in the following, the superscript $(m)$ labels the $m$-th circuit in the sampled batch.
For each sampled circuit, we prepare the state $\ket{\Psi(s^{(m)})}$ on the quantum device and measure it $N_{\mathrm{shot}}$ times in the computational basis. The resulting bit strings are first filtered so that only Slater determinants in the target particle-number sector are retained. We then apply the symmetry-completion procedure introduced by Sugisaki et al.~\cite{Sugisaki2025HSBQSCI}, which closes the sampled open-shell determinants under the minimal set of $\alpha/\beta$ permutations needed for the completed determinant space to support spin eigenfunctions. The explicit construction is given in Appendix~\ref{app:symmetry_completion}. Applying the QSCI procedure to the completed determinant pool yields a truncated subspace $S_{d_{\max}}^{(m)}$, from which classical diagonalization gives the energy $E^{(m)}$. We then define the reward for the $m$-th circuit as
\begin{equation}
    r^{(m)}:=-E^{(m)}.
\end{equation}

After the rewards have been evaluated, we keep the sampled batch fixed and perform multiple policy updates on the same data, following the Group Relative Policy Optimization (GRPO) scheme~\cite{Shao2024DeepSeekMath}. In this work, we use 30 policy updates for each sampled batch. For this fixed batch, we first normalize the rewards within the batch as
\begin{equation}
    \bar r:=\frac{1}{M}\sum_{m=1}^{M} r^{(m)},
    \qquad
    \sigma_r:=\sqrt{\frac{1}{M}\sum_{m=1}^{M}\left(r^{(m)}-\bar r\right)^2},
\end{equation}
and define the corresponding advantage by
\begin{equation}
    \hat A^{(m)}:=\frac{r^{(m)}-\bar r}{\sigma_r}.
\end{equation}
We then update the policy by minimizing the GRPO loss function:
\begin{equation}
    \begin{aligned}
    \mathcal L_{\mathrm{GRPO}}(\theta)
    &:=
    -\frac{1}{M}\sum_{m=1}^{M}\frac{1}{L}\sum_{t=1}^{L}
    \min\Bigl(
    \rho_t^{(m)}(\theta)\hat A^{(m)}, \\
    &\qquad\qquad
    \mathrm{clip}\!\left(\rho_t^{(m)}(\theta),1-\epsilon,1+\epsilon\right)\hat A^{(m)}
    \Bigr),
    \end{aligned}
\end{equation}
where $\epsilon$ is the clipping parameter, set to $0.2$ in this work, and $\rho_t^{(m)}(\theta)$ is the importance ratio defined by
\begin{equation}
    \rho_t^{(m)}(\theta):=
    \frac{\pi_\theta\!\left(s_t^{(m)}\middle|q,s_{<t}^{(m)}\right)}
    {\pi_{\theta_{\mathrm{old}}}\!\left(s_t^{(m)}\middle|q,s_{<t}^{(m)}\right)}.
\end{equation}
Here, $\pi_{\theta_{\mathrm{old}}}$ denotes the frozen policy used to sample the circuits in the current batch, and it is kept fixed during the subsequent policy updates on that batch.

Since each iteration evaluates a batch of $M$ sampled circuits, the quantum-circuit measurement and the subsequent classical diagonalization are carried out a total of $M\times N_{\mathrm{iter}}$ times over the full optimization process.

\subsection{Refinement workflow}\label{sec:refinement}

\begin{figure}[t]
    \centering
    \includegraphics[width=\columnwidth,page=1]{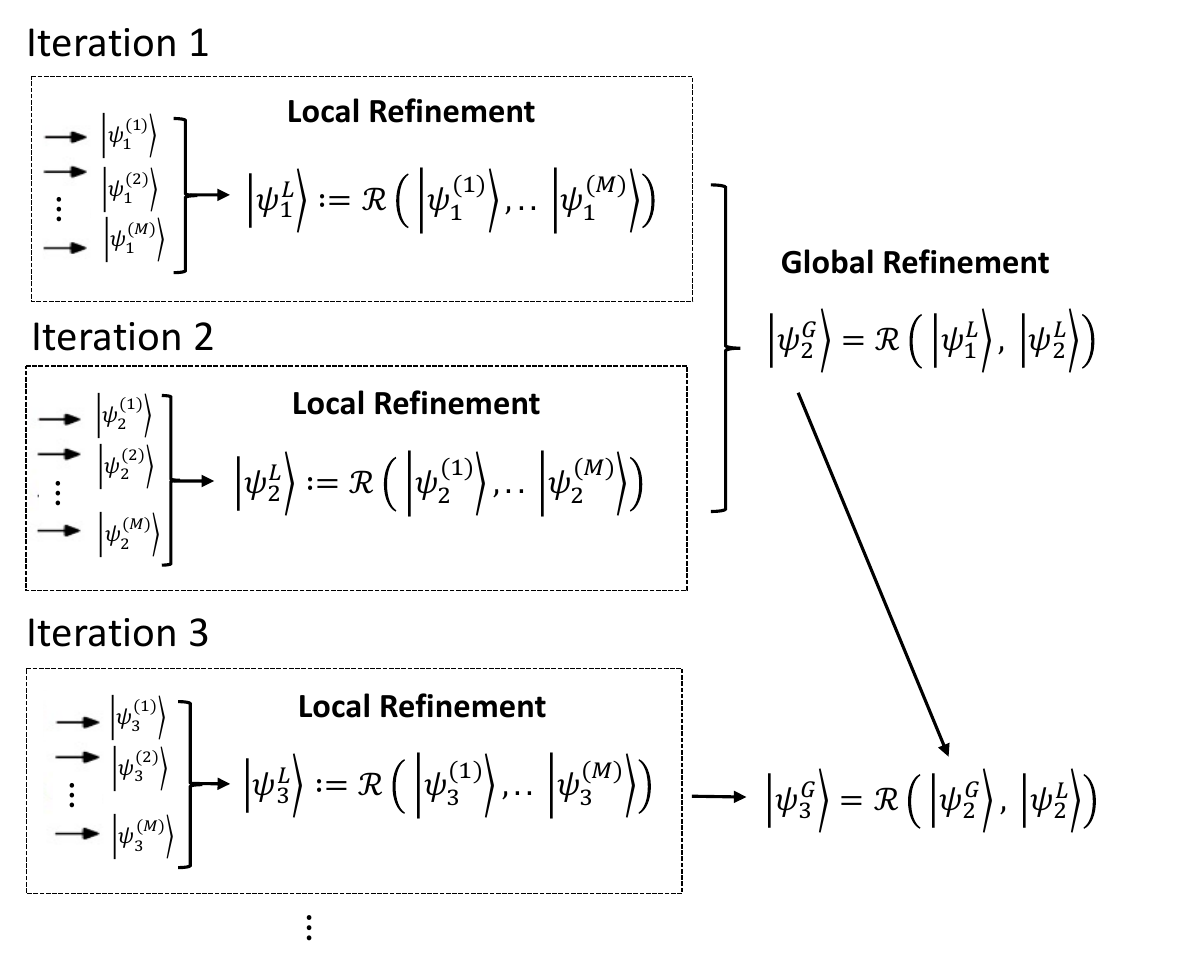}
    \caption{\textbf{Refinement workflow.} At iteration $i$, the local refinement merges the $M$ wavefunctions obtained from the current batch of circuits into a single sparse wavefunction $\ket{\psi_i^L}$. Global refinement then recursively combines $\ket{\psi_i^L}$ with the refined wavefunction carried over from the previous iteration, $\ket{\psi_{i-1}^G}$, to produce $\ket{\psi_i^G}$.}
    \label{fig:refinement}
\end{figure}

The diverse quantum circuits generated during the optimization process also yield approximate wavefunctions, which encode information about significant determinants. We therefore introduce a refinement map $\mathcal R$ that merges several wavefunctions into a single sparse wavefunction with at most $d_{\max}$ determinants.

Consider $n$ input wavefunctions,
\begin{equation}
    \ket{\psi^{(j)}}=\sum_{x\in \mathcal D^{(j)}} a_x^{(j)}\ket{x},
    \qquad j=1,\ldots,n,
\end{equation}
where $a_x^{(j)}=0$ is understood for $x\notin \mathcal D^{(j)}$. In the non--orthogonal basis formed by these wavefunctions, we solve the generalized eigenvalue problem
\begin{equation}
    \begin{aligned}
        \mathbf H\mathbf c &= E\mathbf S\mathbf c, \\
        H_{jk} &:= \matrixel{\psi^{(j)}}{\hat H}{\psi^{(k)}}, \\
        S_{jk} &:= \innerproduct{\psi^{(j)}}{\psi^{(k)}}.
    \end{aligned}
\end{equation}
Let $\mathbf c=(c_1,\ldots,c_n)$ be the eigenvector associated with the lowest eigenvalue. This defines the mixed wavefunction
\begin{equation}
    \ket{\psi_{\mathrm{mix}}}:=\sum_{j=1}^{n} c_j\ket{\psi^{(j)}}
    =\sum_{x\in \mathcal D_{\cup}} A_x\ket{x},
\end{equation}
with
\begin{equation}
    \mathcal D_{\cup}:=\bigcup_{j=1}^{n}\mathcal D^{(j)},
    \qquad
    A_x:=\sum_{j=1}^{n} c_j a_x^{(j)}.
\end{equation}
We then identify the $d_{\max}$ determinants with the largest weights $|A_x|^2$ and use them to define the refined determinant set
\begin{equation}
    \mathcal D_{\mathrm{refined}}
    :=
    \mathrm{Top}_{d_{\max}}\!\left(\{|A_x|^2\}_{x\in \mathcal D_{\cup}}\right).
\end{equation}
The selected determinant set is then used to define a refined subspace, and classical diagonalization of the Hamiltonian in this subspace yields the refined wavefunction,
\begin{equation}
    \ket{\psi_{\mathrm{refined}}}
    :=
    \mathcal R\!\left(\ket{\psi^{(1)}},\ldots,\ket{\psi^{(n)}}\right).
\end{equation}

We now apply this map to the QSCI states obtained during the optimization loop. At iteration $i$, let $\ket{\psi_i^{(m)}}$ be the approximate QSCI ground state obtained from the $m$th circuit in the current batch. As illustrated in Figure~\ref{fig:refinement}, the local refinement combines these $M$ states into
\begin{equation}
    \ket{\psi_i^L}:=\mathcal R\!\left(\ket{\psi_i^{(1)}},\ldots,\ket{\psi_i^{(M)}}\right).
\end{equation}
The global refinement then propagates this information across iterations by recursively combining the current local refinement with the accumulated result from the previous iteration:
\begin{equation}
    \begin{aligned}
        \ket{\psi_1^G} &:= \ket{\psi_1^L}, \\
        \ket{\psi_i^G} &:= \mathcal R\!\left(\ket{\psi_{i-1}^G},\ket{\psi_i^L}\right)
        \quad (i\ge 2).
    \end{aligned}
\end{equation}
Thus, the local refinement removes redundancy among circuits evaluated in the same batch, whereas the global refinement preserves determinants that were found to be significant in earlier iterations even when they are not resampled later. Since both $\ket{\psi_i^L}$ and $\ket{\psi_i^G}$ contain at most $d_{\max}$ determinants, their variational energies can be evaluated entirely classically with the same determinant budget.

\section{Numerical results}
\label{sec:result}
In this section, we perform numerical simulations to assess the proposed framework, focusing on comparisons with existing QSCI algorithms and scaling with system size. To compare algorithmic performance, we consider N$_2$ in the STO-3G~\cite{Hehre1969STO3G} basis with the $(10e,8o)$ active space at $R_{\mathrm{NN}}=1.1$, 1.8, and 2.5~\AA{}, and evaluate sampling efficiency, gate efficiency, and wavefunction compactness. Here, $(N_e,M_o)$ denotes an active space consisting of $N_e$ electrons and $M_o$ molecular orbitals. To examine system-size scaling, we consider N$_2$ in the 6-31G~\cite{Hehre1972_631G} basis at $R_{\mathrm{NN}}=1.8$~\AA{} with the $(10e,8o)$, $(10e,14o)$, and $(10e,16o)$ active spaces, corresponding to 16, 28, and 32 qubits, respectively. All simulations in this section are performed on classical computers. Molecular integrals are obtained from restricted Hartree--Fock calculations with PySCF~\cite{Sun2018PySCF}, and the active orbitals are selected from the resulting canonical orbitals in energy order around the HOMO--LUMO window. Unless otherwise stated, the proposed ansatz optimization uses $M=10$ circuits per iteration, $N_{\mathrm{shot}}=10^5$ measurements per circuit, and $N_{\mathrm{iter}}=100$ iterations. The policy model is a decoder-only GPT-2 model~\cite{Radford2019GPT2} implemented with the Hugging Face \texttt{transformers} library~\cite{Wolf2020Transformers}; the policy parameters are updated by GRPO using AdamW~\cite{LoshchilovHutter2017AdamW} with learning rate $5\times10^{-6}$ and weight decay $0.01$. During autoregressive generation, a repetition penalty of $1.2$ is applied. Quantum-circuit simulations were carried out with CUDA-Q~\cite{cudaq}, and the subspace Hamiltonian construction and diagonalization were performed with PyCI~\cite{Richer2024PyCI}, with all computations run on an NVIDIA H100 system in the ABCI-Q environment.

As benchmark baselines, we compare against the local unitary cluster Jastrow (LUCJ) ansatz~\cite{Matsuzawa2020Jastrow,Motta2023LUCJ}, time-evolved QSCI~\cite{Mikkelsen2025TEQSCI,Sugisaki2025HSBQSCI,Yu2025SKQDv3}, SqDRIFT~\cite{Piccinelli2025SqDRIFT}, and an exact ground-state sampling reference. For the time-evolved QSCI baselines, the input state is
\begin{equation}
    \ket{\Psi_{k,\Delta t}} = e^{-i\hat H k\Delta t}\ket{\Phi_{\mathrm{HF}}}.
\end{equation}
In this work, we consider two variants based on this form: a single time-evolved variant and a multiple time-evolved variant. For the single time-evolved variant, we fix $k=1$ and choose $\Delta t$ separately for each molecule near the empirically optimal value. For the multiple time-evolved variant, we fix $\Delta t=1$ and use $k=1,2,3,4,5$, splitting the shots uniformly among the five states. In both variants, the real-time evolution circuit is constructed by first-order Trotterization with a single Trotter step. For SqDRIFT, we fix the number of randomizations to 500 and use $k=1,2,3$, while the number of sampled excitations is chosen separately for each experiment. The parameter settings of these baselines are chosen using standard choices together with empirical sweeps so as to provide strong performance within the circuit-depth regime considered here. While further fine-tuning may slightly change the quantitative results, it is not expected to alter the qualitative trends discussed below. As an additional reference, we also sample determinants from the exact complete active space configuration interaction (CASCI) ground-state distribution within the corresponding active space. Unless otherwise stated, all sampled determinants are post-processed by the same symmetry-completion and classical diagonalization procedure described in Sec.~\ref{sec:methods}.

\subsection{Optimization performance}
To first verify that the GQE-based optimization improves the QSCI input state in practice, we consider N$_2$ in STO-3G with the $(10e,8o)$ active space at the stretched geometry $R_{\mathrm{NN}}=2.5$~\AA{}. Here, we deliberately adopt a very shallow circuit with only $L=10$ operators to make the optimization problem demanding. For comparison to a random baseline, we also generate $M$ circuits by selecting $L=10$ operators uniformly at random from the operator pool and apply the same QSCI post-processing to them. 

\begin{figure}[htbp]
    \centering
    \begin{subfigure}[t]{\linewidth}
        \centering
        \includegraphics[width=0.95\linewidth]{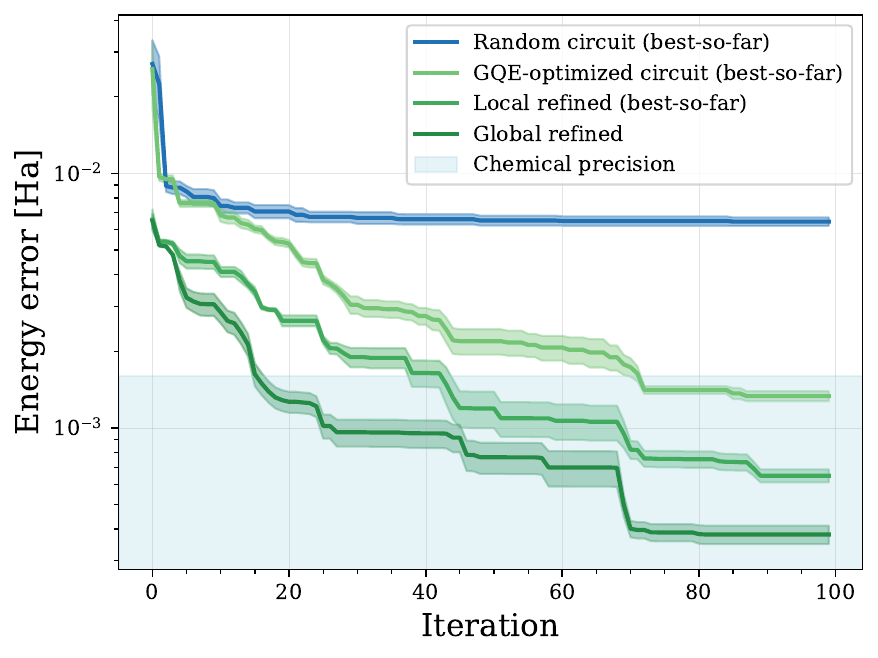}
    \end{subfigure}
    
    \vspace{0.5em}
    \begin{subfigure}[t]{\linewidth}
        \centering
        \includegraphics[width=0.95\linewidth]{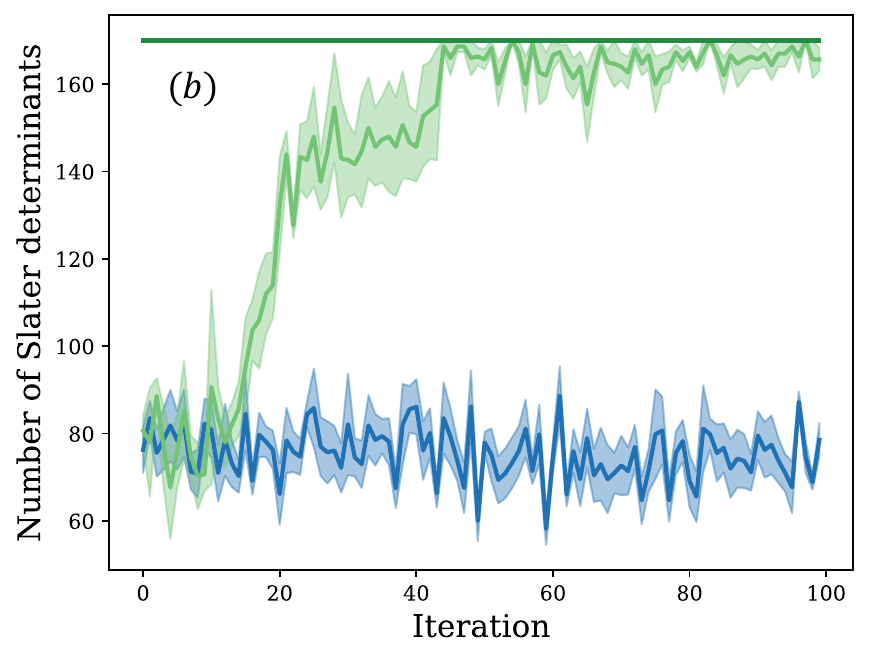}
    \end{subfigure}
    \caption{\textbf{Optimization history for \boldmath$\mathrm{N}_2 (10e,8o)$\unboldmath\ with \boldmath$d_{\max}=170$\unboldmath.} Panel (a) shows the best-so-far energy error up to each iteration, while panel (b) shows the number of retained determinants after symmetry completion and truncation. In both panels, the solid curves denote the mean over five independent optimization runs and the shaded bands denote one standard deviation. The horizontal band in panel (a) indicates chemical precision, conventionally taken as 1 kcal/mol
    ($\approx 1.5936 \times 10^{-3}$ Ha), relative to the CASCI energy.}
    \label{fig:optimization_history}
\end{figure}

Figure~\ref{fig:optimization_history} summarizes the first 100 optimization iterations. The energy error relative to the CASCI energy in Figure~\ref{fig:optimization_history}(a) is reported as the best-so-far value up to each iteration, averaged over five independent runs. While the random baseline rapidly plateaus and shows no meaningful improvement thereafter, the GQE-optimized circuit continues to achieve lower energy and eventually reaches the threshold for chemical precision. The refinement steps are even more effective: the local refinement crosses the chemical precision threshold earlier, and the global refinement yields the lowest final error among all variants. If the circuit optimization were perfect, the final optimized and locally refined energies would approach the globally refined energy. Instead, the globally refined curve remains clearly lower even at the end of the run. This suggests that determinants discovered transiently during the early and intermediate stages still contribute substantially to the final energy and that retaining them stabilizes the optimization.

Figure~\ref{fig:optimization_history}(b) shows the number of Slater determinants retained after the post-processing step for the best circuit found so far. The random circuits keep returning only $\sim 70$--$90$ determinants with $10^5$ shots, far below the cap $d_{\max}=170$, which indicates severe resampling of already-seen configurations. The GQE-optimized circuits, by contrast, steadily expand the determinant set and saturate the budget after roughly 45 iterations. Notably, the energy continues to improve even after this saturation, especially for the global refinement. Hence the late-stage improvement is not due to a larger subspace, but to a better composition of a subspace of essentially fixed size. 

\begin{figure*}[htbp]
    \centering
    \includegraphics[width=\textwidth,page=1]{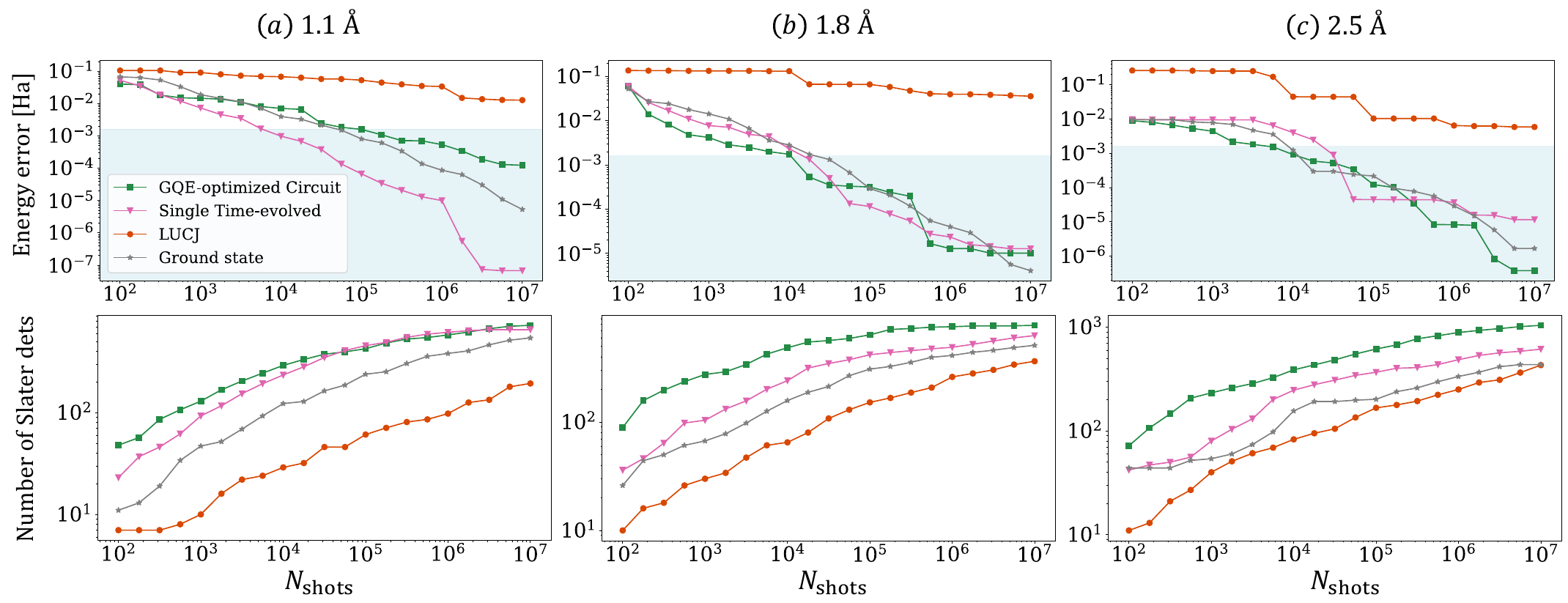}
    \caption{\textbf{Sampling efficiency for \boldmath$\mathrm{N}_2 (10e,8o)$\unboldmath.} The top row shows the energy error and the bottom row the number of unique determinants as functions of the total number of shots. The optimized ansatz is the best circuit obtained after $N_{\mathrm{iter}}=100$ optimization iterations with $L=140$, chosen so that its two-qubit gate count is close to that of a three-layer LUCJ circuit (see Table~\ref{tab:sampling_gate_counts}). For the single time-evolved baseline, we fix $\Delta t=0.8$, which was found to be near-optimal across all three bond lengths in a sweep over $\Delta t=0.1,0.2,\ldots,3.0$. The exact-ground-state reference samples directly from the CASCI distribution. The blue band indicates chemical precision.}
    \label{fig:sample_efficiency}
\end{figure*}

Taken together, Figure~\ref{fig:optimization_history} shows that the proposed workflow succeeds through two complementary mechanisms: the circuit optimization preferentially discovers determinants that are significant for QSCI, while the refinement procedures accumulate them into a more accurate subspace. We further note that, in additional experiments, this GQE-based discrete optimization was also found to remain more stable than a VQE-based continuous-parameter optimization when the determinant budget $d_{\max}$ is tightly constrained. A detailed comparison is provided in Appendix~\ref{app:vqe_comparison}.

\subsection{Sampling efficiency}
Recent work has emphasized that QSCI fundamentally faces a sampling-efficiency challenge~\cite{Reinholdt2025FatalFlaw}. Unlike classical selected-CI algorithms such as HCI~\cite{Holmes2016HCI,Sharma2017SHCI}, which expand the determinant space through an explicit tree-like search, QSCI acquires configurations by measurement and therefore inevitably risks resampling the same determinants many times. When the sampling distribution is too strongly concentrated on a few determinants, the number of shots needed to recover the set of configurations required for an accurate subspace increases. This problem has been mitigated by Cartesian-product expansions of the sampled determinant set in past work~\cite{RobledoMoreno2025ScienceAdv}, but such expansions can easily sacrifice compactness by introducing many determinants that are not significant. In the present work, we therefore keep the expansion step minimal, restricting it to symmetry completion, and instead investigate to what extent ansatz optimization can improve the sampling efficiency itself.

In this section, we compare the sampling efficiency of the GQE-optimized ansatz with other QSCI baselines. For the optimized ansatz, we select the best circuit obtained after 100 optimization iterations and fix the circuit length to $L=140$. This value was chosen so that the optimized circuit has nearly the same two-qubit gate count as a three-layer LUCJ circuit; the corresponding gate counts are summarized in Table~\ref{tab:sampling_gate_counts}. For the single time-evolved baseline, we fix $\Delta t=0.8$, which was found to be close to optimal at all three bond lengths in a sweep over $\Delta t=0.1,0.2,\ldots,3.0$. Figure~\ref{fig:sample_efficiency} shows the energy error (top row) and the number of unique sampled determinants (bottom row) as functions of the total number of shots.

The advantage over LUCJ is clear at all bond lengths. Even when the shot count is increased to $10^7$, LUCJ yields only a relatively small determinant set and remains above chemical precision relative to the CASCI energy. The optimized circuit, by contrast, produces hundreds of determinants at moderate shot counts and yields a lower energy error. Importantly, this improvement is obtained with essentially the same two-qubit gate count as the three-layer LUCJ circuit but with a smaller total gate count, as shown in Table~\ref{tab:sampling_gate_counts}. Thus, task-specific ansatz optimization dramatically improves the sampling efficiency of QSCI at comparable or lower circuit cost.

\begin{figure*}[htbp]
    \centering
    \includegraphics[width=\textwidth,page=1]{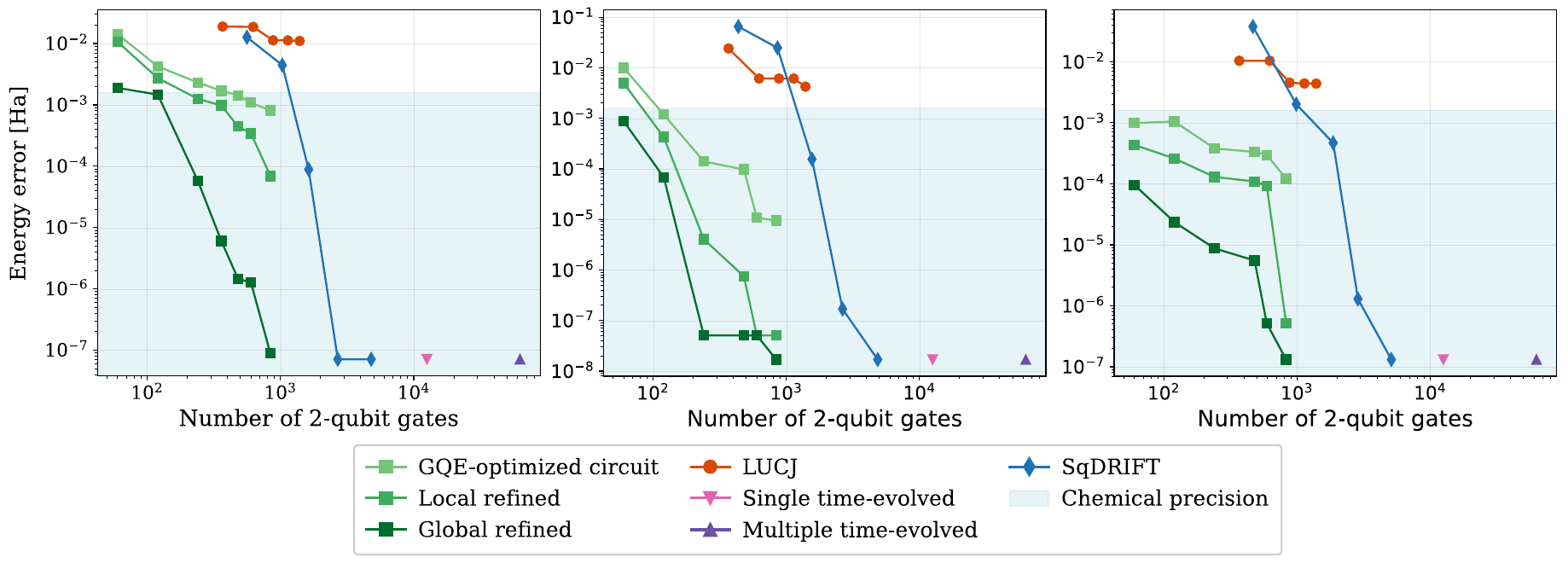}
    \caption{\textbf{Two-qubit-gate efficiency for \boldmath$\mathrm{N}_2 (10e,8o)$\unboldmath.} Energy error is plotted against the number of two-qubit gates for $R_{\mathrm{NN}}=(a)\ 1.1$~\AA{}, $(b)\ 1.8$~\AA{}, and $(c)\ 2.5$~\AA{}. The GQE-optimized circuit is scanned by varying the circuit length from $L=10$ to $140$, and the plotted gate counts are the maximum two-qubit-gate counts among the circuits generated during the corresponding optimization runs. LUCJ uses 1 to 5 layers, and SqDRIFT uses 5 to 200 sampled excitations with 500 randomizations and $k=1,2,3$. The single- and multiple-time-evolved baselines follow the settings defined in Sec.~\ref{sec:methods}. All methods use the same total shot budget of $10^8$ measurements. Local and global refinement inherit the gate count of the underlying optimized circuit because refinement is entirely classical post-processing.}
    \label{fig:gate_efficiency}
\end{figure*}

The comparison with the single time-evolved state depends strongly on bond length, or equivalently on how accurately the Hartree--Fock reference already approximates the ground state. At $R_{\mathrm{NN}}=1.1$~\AA{}, the single time-evolved baseline is the most sample-efficient among all states considered here; although Table~\ref{tab:sampling_gate_counts} shows that it is also more than an order of magnitude more expensive in gate count than the optimized circuit. As the bond is stretched, however, the optimized circuit begins to outperform the baseline. At $1.8$~\AA{} the two are already comparable, while at $2.5$~\AA{} the optimized circuit yields both more determinants and a lower final energy. A natural explanation is the decreasing overlap between the Hartree--Fock determinant and the exact ground state, which drops from $0.957$ to $0.652$ to $0.314$ as $R_{\mathrm{NN}}$ increases from $1.1$ to $1.8$ to $2.5$~\AA{}. When the Hartree--Fock state is already close to the ground state, a real-time evolution can efficiently expose the necessary determinants. In the strongly stretched regime, the same strategy becomes less well aligned with the target subspace, whereas the optimized circuit can reshape the sampling distribution more directly. 

It is also worth noting that the exact ground state is not automatically the most sample-efficient input state for QSCI. Depending on the shot budget and bond length, direct sampling from the exact distribution can yield fewer useful determinants than the optimized or time-evolved states. This again highlights that QSCI benefits not merely from high overlap with the ground state, but from a sampling distribution that preferentially exposes the determinants most useful for constructing an accurate subspace.

\begin{table}[htbp]
    \centering
    \caption{Gate counts for the circuits used in Figure~\ref{fig:sample_efficiency}. Gate counts are evaluated assuming all-to-all connectivity and a standard decomposition into arbitrary-angle single-qubit rotation gates, CX gates, and single-qubit Clifford gates.}
    \begin{tabular}{lcc}
        \hline
        Circuit & 2-qubit gates & Total gates \\
        \hline
        GQE-optimized circuit ($L=140$) & 834 & 3031 \\
        LUCJ (3 layers) & 880 & 5850 \\
        Single time-evolved & 12544 & 51186 \\
        \hline
    \end{tabular}
    \label{tab:sampling_gate_counts}
\end{table}

\subsection{Gate efficiency}
The gate cost is one of the central concerns in practical implementations of QSCI on quantum hardware. In the previous subsection, we showed that ansatz optimization can achieve high sampling efficiency at a gate count comparable to that of LUCJ. Here we broaden the scan over circuit size and compare how many entangling gates are required to reach chemical precision across different QSCI input-state families. The GQE-optimized ansatz is scanned by varying the circuit length from $L=10$ to $140$ with $d_{\max}=\infty$, LUCJ by varying the number of layers from 1 to 5, and SqDRIFT by varying the number of sampled excitations from 5 to 200 with 500 randomizations and $k=1,2,3$. To make the comparison fair, all QSCI baselines are given the same total shot budget as used by the optimization itself, namely $N_{\mathrm{shot}} \times M \times N_{\mathrm{iter}}=10^8$ measurements. Because local and global refinement are purely classical post-processing steps, their curves are plotted at the same gate count as the underlying optimized circuit. A complementary comparison using the number of rotation gates as a fault-tolerant cost proxy is given in Appendix~\ref{app:rotation_gate_efficiency}.

\begin{figure*}[t]
    \centering
    \includegraphics[width=\textwidth,page=1]{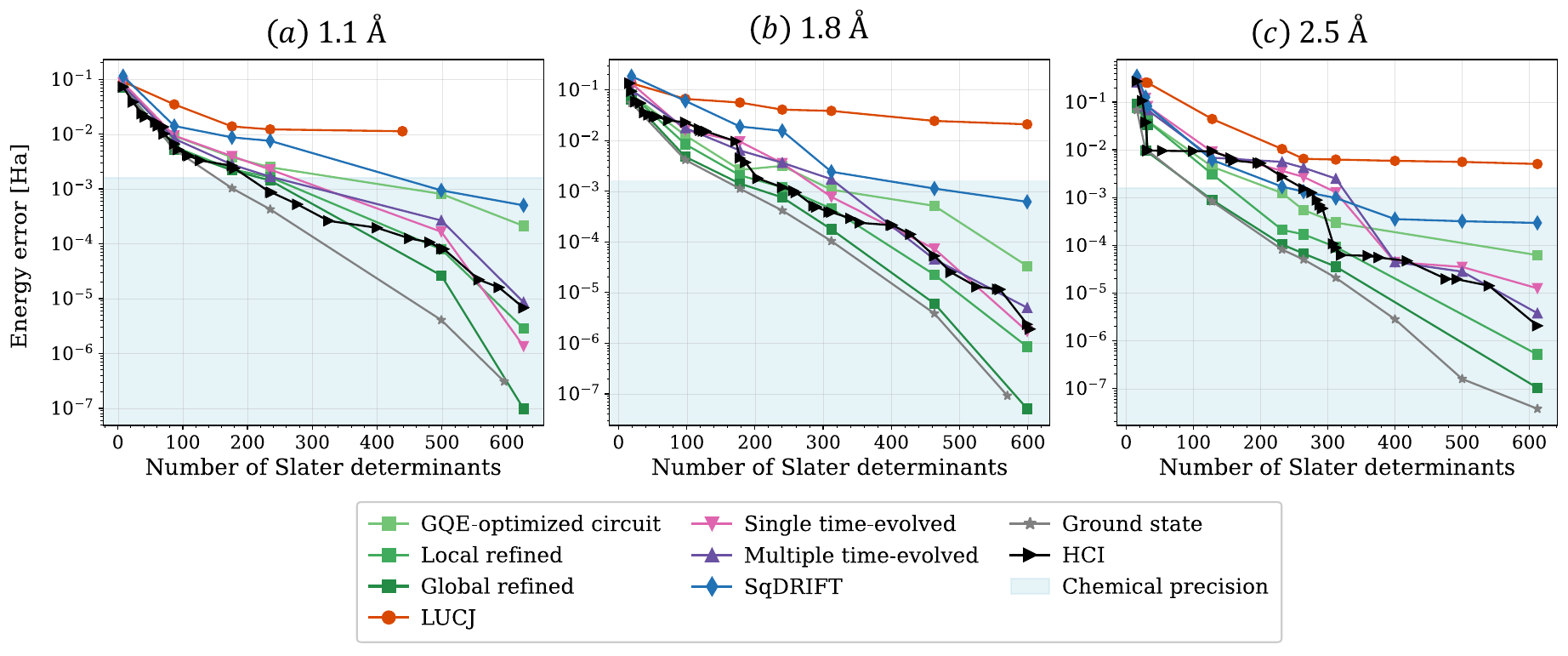}
    \caption{\textbf{Wavefunction compactness for \boldmath$\mathrm{N}_2 (10e,8o)$\unboldmath.} Energy error is plotted against the number of determinants in the final QSCI subspace. The optimized ansatz is the best circuit obtained after $N_{\mathrm{iter}}=100$ iterations with $L=140$. LUCJ is fixed to 3 layers, and SqDRIFT uses 100 sampled excitations. All quantum baselines use the same total shot budget of $10^8$ measurements. The HCI curve is obtained by varying the selection threshold. The exact-ground-state reference samples directly from the CASCI distribution. The blue band indicates chemical precision.}
    \label{fig:compactness}
\end{figure*}

On noisy intermediate-scale quantum (NISQ) hardware, the number of two-qubit entangling gates is a particularly important cost proxy, because such gates typically dominate both error accumulation and executable circuit depth. As shown in the top row of Figure~\ref{fig:gate_efficiency}, the GQE-optimized circuits consistently achieve a given accuracy at much lower two-qubit-gate cost than the time-evolved baselines across all three bond lengths. Averaged over the three geometries, the two-qubit gate count required to reach chemical precision is reduced by 98\% relative to the single-step first-order Trotterized approximation and by 83\% relative to SqDRIFT. Although these time-evolved baselines can eventually become highly accurate, they do so only at substantially larger gate counts. In contrast, the optimized circuits enter the chemical-precision window with fewer than $10^3$ two-qubit gates, after which the refined subspaces reduce the error still further. LUCJ remains above chemical precision throughout the scanned range even at five layers. These results show that the learned ansatz improves not only sampling efficiency but also the entangling-gate efficiency of QSCI state preparation.

\subsection{Wavefunction compactness}

We now turn to the classical side of the cost. QSCI is more useful when it reaches a target accuracy with fewer determinants, because the diagonalization cost grows with the subspace size. In Figure~\ref{fig:compactness}, we therefore plot the energy error against the number of determinants used for classical diagonalization. For the GQE-optimized circuit results, each point is taken from the best circuit obtained after 100 optimization iterations with $L=140$ in an independent optimization run performed with the corresponding value of $d_{\max}$. LUCJ is fixed to 3 layers, SqDRIFT uses 100 sampled excitations, and all quantum baselines are given the same total shot budget of $10^8$.

The first comparison is between the optimized circuit and the 3-layer LUCJ circuit, which has a similar circuit cost. At all three geometries, the optimized circuit gives substantially lower errors at the same subspace size, showing that the ansatz learned by GQE already produces a much more compact and accurate wavefunction before any refinement is applied. This contrast is particularly clear at $R_{\mathrm{NN}}=1.1$~\AA{}, where the LUCJ curve terminates at a relatively small number of determinants and does not even reach the larger-subspace region covered by the other methods. This behavior is consistent with the sampling-efficiency result in Figure~\ref{fig:sample_efficiency}: even under the same measurement budget, the LUCJ state fails to provide enough distinct significant determinants, and the resulting compactness is limited from the outset.

The effect of refinement is then even more pronounced. The local refinement makes the subspace markedly denser in significant determinants and reaches a compactness comparable to, or better than, the time-evolved states across the three bond lengths. Global refinement improves this further and becomes competitive with HCI, with the balance depending on bond length. At $R_{\mathrm{NN}}=1.1$~\AA{}, the globally refined subspace already gives a lower final error at around 600 determinants, but the subspace size required to reach chemical precision is still slightly smaller for HCI. At $R_{\mathrm{NN}}=1.8$~\AA{}, this balance reverses, and the globally refined subspace is more compact over most of the plotted range, reaching a given accuracy with fewer determinants. This trend is most clearly seen at $R_{\mathrm{NN}}=2.5$~\AA{}, where the advantage of the global refinement becomes pronounced: chemical precision is reached with a subspace about 50\% smaller than that required by HCI.

A plausible explanation is that strong static correlation makes the significant configurations less accessible to the hierarchical tree search used in HCI~\cite{Holmes2016HCI,Sharma2017SHCI}. In these methods, the determinant space is expanded step by step from a small reference space such as the Hartree--Fock determinant, with new configurations generated according to the connectivity defined by the Slater--Condon rules. Because each determinant is then connected to only a limited number of nearby configurations, the search advances through a sparse connectivity graph. As a result, when significant determinants lie several rearrangements away from the dominant reference configurations, they may be incorporated only after many intermediate candidates have already been explored. In contrast, the optimized circuit can discover such determinants more globally through sampling, and the global refinement can retain and accumulate them across the optimization trajectory. This difference is modest near equilibrium, becomes clear at $1.8$~\AA{}, and is most pronounced at $2.5$~\AA{}, where determinants beyond the Hartree--Fock configuration become dominant.

\subsection{System-size scaling with fixed shot count}
Finally, we test the scaling behavior of this workflow. Although the present benchmarks are limited to small molecules, QSCI is motivated by much larger molecules, and recent SQD calculations have already reached 77-qubits~\cite{RobledoMoreno2025ScienceAdv}. In that regime, one of the main practical bottlenecks is the measurement budget. As the many-electron Hilbert space grows combinatorially with the number of orbitals, the determinant subspace required to achieve a given accuracy can also grow substantially. At the same time, it is not realistic in practice to compensate for this growth simply by increasing the shot count to $10^9$ or $10^{10}$. Therefore, we aim to determine how effective the present optimization and refinement workflow remains when the measurement budget is fixed. To examine this point, we consider N$_2$ in the 6-31G basis at $R_{\mathrm{NN}}=1.8$~\AA{} and enlarge the active space to $(10e,14o)$ and $(10e,16o)$, corresponding to 28 and 32 qubits. We keep the per-circuit shot budget fixed at $N_{\mathrm{shot}}=10^5$, with $M=10$ circuits and $N_{\mathrm{iter}}=100$. For these larger calculations we use circuit lengths $L=600$ and $L=800$ for the 28- and 32-qubit cases, respectively.

\begin{figure}[htbp]
    \centering
    \begin{subfigure}[t]{\linewidth}
        \centering
        \includegraphics[width=0.95\linewidth]{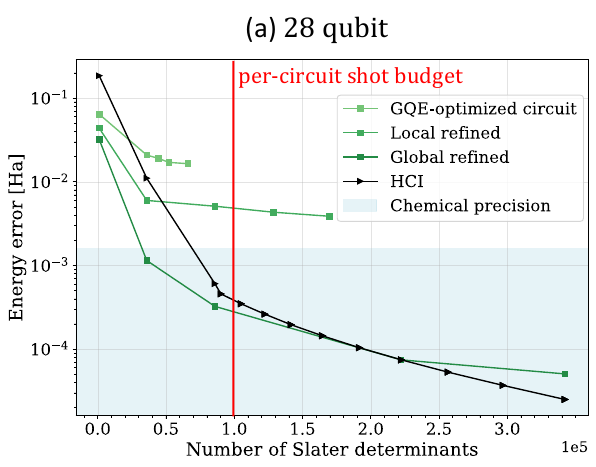}
    \end{subfigure}
    
    \vspace{0.5em}
    \begin{subfigure}[t]{\linewidth}
        \centering
        \includegraphics[width=0.92\linewidth]{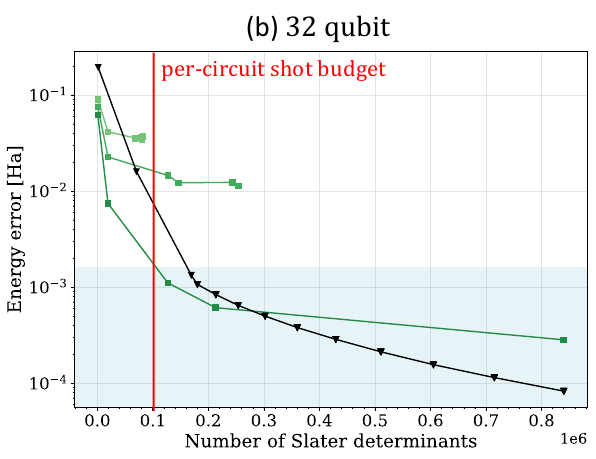}
    \end{subfigure}
    \caption{\textbf{Scaling with active-space size at fixed per-circuit shot count.} The two panels show the larger active spaces, (a) $(10e,14o)$ (28 qubits) and (b) $(10e,16o)$ (32 qubits). For these larger calculations we use $N_{\mathrm{iter}}=100$, with circuit lengths $L=600$ and $L=800$ for the 28- and 32-qubit cases, respectively. The red vertical line indicates the per-circuit shot budget $N_{\mathrm{shot}}=10^5$.}
    \label{fig:scaling}
\end{figure}

Figure~\ref{fig:scaling} shows a clear hierarchy among the three variants of the proposed workflow. In both the 28- and 32-qubit cases, the GQE-optimized circuit alone remains far from chemical precision, indicating that under the present fixed-shot setting the determinant set obtained from a single optimized circuit is not sufficient by itself. The local refinement improves the results, but the most substantial gain comes from the global refinement, which accumulates determinant information across circuits and optimization iterations. As a result, the globally refined subspace recovers enough significant determinants to reach chemical precision in both active spaces. Notably, even in the 32-qubit case, chemical precision is reached at a smaller subspace size than in HCI, showing that within this regime the proposed workflow can still construct a more efficient subspace.

This behavior suggests a practical criterion for the useful regime of the method under fixed shots. When chemical precision can be achieved with a subspace dimension on the order of $N_{\mathrm{shot}}$, the present workflow retains an advantage over HCI. In this sense, the main message of Figure~\ref{fig:scaling} is that the proposed workflow remains effective under fixed shots as long as the target-accuracy regime is accessible within a subspace size comparable to the available number of samples.

\section{Conclusion and Discussion}
\label{conclusion}
In this work, we developed a GQE-based ansatz-optimization framework for quantum-selected configuration interaction (QSCI) aimed at constructing compact and accurate wavefunctions within tight gate-cost budgets. The proposed workflow trains a Transformer policy using the QSCI subspace energy as the reward and complements this search with the local and the global refinement procedures that aggregate determinant information generated throughout the optimization. Together, these components enable low-cost circuit search and progressive reconstruction of compact and accurate QSCI wavefunctions.

Numerical simulations on N$_2$ show that this strategy can construct accurate and compact QSCI subspaces at substantially lower circuit cost. Relative to fixed LUCJ circuits~\cite{Matsuzawa2020Jastrow,Motta2023LUCJ}, the optimized ansatz achieves markedly lower energy errors at comparable gate counts. Compared with time-evolved baselines~\cite{Mikkelsen2025TEQSCI,Sugisaki2025HSBQSCI,Yu2025SKQDv3,Piccinelli2025SqDRIFT}, it reaches the same accuracy regime with far fewer entangling and rotation gates, indicating a more favorable balance between accuracy and circuit cost. Moreover, local refinement reaches compactness comparable to or better than the time-evolved baselines, and the global refinement produces subspaces competitive with, and for stretched-bond N$_2$ even smaller than, HCI~\cite{Holmes2016HCI,Sharma2017SHCI} at the same target accuracy. These results show that circuit optimization and refinement play complementary roles: the former lowers the gate cost required to generate useful determinant distributions, while the latter converts them into compact and accurate QSCI wavefunctions.

Several challenges remain. A central one is the optimization cost of the present workflow. A full training run requires $M\times N_{\mathrm{iter}}$ circuit evaluations, each accompanied by $N_{\mathrm{shot}}$ measurements and a classical diagonalization. Additionally, this overhead grows with system size as both the search space and the determinant budget increase. A natural direction is therefore large-scale pre-training on diverse molecular systems, followed by transfer learning to new geometries and related molecules, in line with future directions envisioned for GQE~\cite{Nakaji2025GQE}. Reusing a learned search bias may reduce the number of task-specific optimization iterations and improve the practicality and scalability of ansatz optimization. Such a reduction in optimization cost would also make experimental validation on quantum hardware more feasible. The shallower circuits found here may offer an advantage in that setting, but the practical benefit must ultimately be judged under realistic noise.

Overall, the present results establish circuit optimization as a practical route to make QSCI more efficient in its use of both quantum and classical resources, and suggest that learned and transferable circuit-design strategies may be used in extending QSCI to larger chemically relevant systems.

\section{Code availability}
The code used in this study is publicly available at \url{https://github.com/moken20/gqe-for-qsci}.

\section*{Acknowledgments}
A part of this work was performed for Council for Science, Technology and Innovation (CSTI), Cross-ministerial Strategic Innovation Promotion Program
(SIP), “Promoting the application of advanced quantum technology platforms to social issues” (Funding
agency: QST). The results presented in this paper were obtained using the ABCI-Q of AIST G-QuAT. RK would like to express gratitude to Kenji Sugisaki for an insightful discussion.

\bibliography{bibliography}

\appendix
\clearpage
\onecolumngrid
\section{Operator pool}
\label{app:operator_pool}

The operator pool used by the Transformer is a fixed discrete vocabulary of unitary operators. Our construction starts from the UCCSD-derived Pauli-time-evolution pool used in GQE~\cite{Nakaji2025GQE}, but simplifies it in the spirit of the single-Pauli-operator pool adopted in ADAPT-QSCI~\cite{Nakagawa2024ADAPTQSCI}. This simplification is motivated by the QSCI setting: the goal is not to reproduce a chemically motivated unitary ansatz as faithfully as possible, but to reshape the sampling distribution over determinants under a tight gate budget.

Let $\mu$ label a spin-orbital single or double excitation relative to the Hartree--Fock reference $\ket{\Phi_{\mathrm{HF}}}$, with anti-Hermitian generator $\hat\tau_\mu$. For example,
\begin{equation}
    \hat\tau_i^a = a_a^\dagger a_i - a_i^\dagger a_a,
    \qquad
    \hat\tau_{ij}^{ab} = a_a^\dagger a_b^\dagger a_j a_i - a_i^\dagger a_j^\dagger a_b a_a,
\end{equation}
where $i,j$ denote occupied spin orbitals and $a,b$ denote virtual spin orbitals in the Hartree--Fock reference. After the Jordan--Wigner mapping and a first-order Trotter decomposition, each excitation contributes a product of Pauli time evolutions,
\begin{equation}
    e^{\theta_\mu \hat\tau_\mu}
    \approx
    \prod_{\ell=1}^{n_\mu} e^{i\theta_\mu c_{\mu\ell}\hat P_{\mu\ell}},
\end{equation}
where $\hat P_{\mu\ell}$ are Pauli strings and $c_{\mu\ell}\in\mathbb R$. The original GQE pool keeps all such Pauli strings and pairs them with a discrete set of time parameters~\cite{Nakaji2025GQE}.

In the present work we modify this construction in three steps. First, we drop the Jordan--Wigner parity string $\prod Z$ from each Pauli term; schematically,
\begin{equation}
    X_q Z_{q-1}\cdots Z_{p+1} Y_p
    \;\mapsto\;
    X_q Y_p,
\end{equation}
which substantially shortens the entangling-gate decomposition. Second, instead of keeping all Pauli terms produced by a given excitation, we retain only one representative Pauli string $\widetilde P_\mu$ for each excitation channel $\mu$. This one-term extraction makes one token correspond to one chemically motivated excitation channel and keeps the vocabulary size manageable. Third, rather than introducing an additional discrete time parameter as in GQE, we absorb the coefficient of the representative Pauli term together with the corresponding classical CCSD amplitude into a fixed scalar $\vartheta_\mu$ and define the token unitary as
\begin{equation}
    U_\mu := e^{i\vartheta_\mu \widetilde P_\mu}.
\end{equation}
The final operator pool is therefore
\begin{equation}
    \mathcal G = \{I\} \cup \{U_\mu\},
\end{equation}
where each $U_\mu$ is associated with a spin-orbital single or double excitation. To keep the pool compact, we include only excitations whose corresponding CCSD amplitudes exceed $10^{-6}$ in magnitude.

This design keeps the optimization problem purely combinatorial: the policy learns only which operators to place and in what order. At the same time, because each token is a single Pauli rotation rather than the full fermionic excitation operator, individual tokens do not exactly conserve particle number or spin projection. In our workflow, measurement outcomes are first post-selected onto the target $(N_\alpha, N_\beta)$ sector, and the resulting determinant set is then processed by the symmetry-completion procedure described in Appendix~\ref{app:symmetry_completion}.

\section{Symmetry-completion}
\label{app:symmetry_completion}

The symmetry-completion step used in this work follows the idea introduced in  Hamiltonian simulation-based QSCI (HSB-QSCI)~\cite{Sugisaki2025HSBQSCI}: before constructing the subspace Hamiltonian, one augments sampled open-shell determinants by adding the missing determinants required to span spin eigenfunctions. The important point is that this step is \emph{not} a direct projection onto a prescribed total-spin irrep such as the singlet sector. Rather, it completes the determinant space within a fixed $(N_\alpha,N_\beta)$ sector so that classical diagonalization of the spin-free electronic Hamiltonian can form the appropriate spin eigenstates.

Let $\widetilde{\mathcal D}$ be the set of unique determinants obtained after filtering the measured bit strings to the target $(N_\alpha,N_\beta)$ sector. For a determinant $\ket{x}$, define the four-valued orbital label
\begin{equation}
    \ell_p(x)
    :=
    \begin{cases}
        2, & (x_p^\uparrow,x_p^\downarrow)=(1,1),\\
        \uparrow, & (x_p^\uparrow,x_p^\downarrow)=(1,0),\\
        \downarrow, & (x_p^\uparrow,x_p^\downarrow)=(0,1),\\
        0, & (x_p^\uparrow,x_p^\downarrow)=(0,0),
    \end{cases}
\end{equation}
and the associated sets of doubly occupied, open-shell, and empty orbitals,
\begin{equation}
    D(x):=\{p\mid \ell_p(x)=2\},
    \qquad
    O(x):=\{p\mid \ell_p(x)\in\{\uparrow,\downarrow\}\},
    \qquad
    V(x):=\{p\mid \ell_p(x)=0\}.
\end{equation}
On the open-shell set we further define
\begin{equation}
    n_\alpha(x):=\sum_{p\in O(x)} x_p^\uparrow,
    \qquad
    n_\beta(x):=\sum_{p\in O(x)} x_p^\downarrow,
\end{equation}
so that $n_\alpha(x)+n_\beta(x)=|O(x)|$.

The symmetry-completion class of $x$ is then defined as
\begin{equation}
    \mathcal C(x)
    :=
    \left\{
    x'\,\middle|\,
    \begin{array}{l}
        D(x')=D(x),\\
        O(x')=O(x),\\
        V(x')=V(x),\\
        n_\alpha(x')=n_\alpha(x)
    \end{array}
    \right\}.
\end{equation}
In words, we keep the spatial occupation pattern fixed and generate all determinants obtained by permuting the $\alpha/\beta$ assignments only on the open-shell orbitals while preserving $N_\alpha$ and $N_\beta$. The completed determinant pool is
\begin{equation}
    \widetilde{\mathcal D}_{\mathrm{SC}}
    :=
    \bigcup_{x\in\widetilde{\mathcal D}} \mathcal C(x).
\end{equation}
For a determinant with $m:=|O(x)|$ open-shell orbitals, the size of its completion class is
\begin{equation}
    |\mathcal C(x)|=\binom{m}{n_\alpha(x)}.
\end{equation}
Closed-shell determinants have $m=0$ and are therefore unchanged.

This formalization matches the verbal prescription in HSB-QSCI~\cite{Sugisaki2025HSBQSCI}. For instance, when a determinant with occupation pattern $(2,\uparrow,\downarrow,0)$ is sampled, the missing partner $(2,\downarrow,\uparrow,0)$ is added as well. The resulting two-determinant space is then able to support spin eigenfunctions, which is exactly the purpose of the completion step.

After symmetry completion, we apply the same determinant-budget truncation as in the main QSCI workflow and retain at most $d_{\max}$ determinants for the classical diagonalization. Because the completion acts only within each sampled spatial occupation pattern, it is substantially more compact than the Cartesian-product expansion used in standard SQD~\cite{RobledoMoreno2025ScienceAdv}.

\section{Supplementary numerical results}
\subsection{Comparing optimization performance with VQE}
\label{app:vqe_comparison}

The main-text results already show that the proposed workflow improves both gate efficiency and wavefunction compactness relative to fixed QSCI input-state families. Those comparisons, however, do not by themselves isolate whether the advantage comes specifically from the GQE-style discrete circuit search or simply from performing some optimization. To clarify this point, we compare the proposed generative quantum eigensolver (GQE)~\cite{Nakaji2025GQE} with a variational quantum eigensolver (VQE)~\cite{Peruzzo2014VQE} baseline under the same QSCI evaluation budget.

\begin{figure}[htbp]
    \centering
    \begin{subfigure}[t]{0.49\linewidth}
        \centering
        \includegraphics[width=\linewidth]{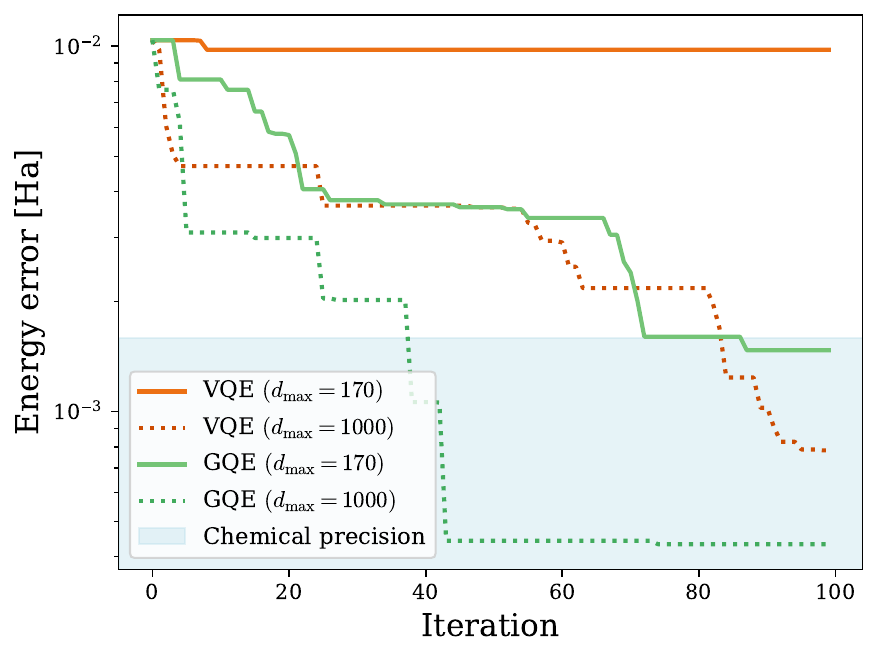}
    \end{subfigure}
    \hfill
    \begin{subfigure}[t]{0.49\linewidth}
        \centering
        \includegraphics[width=\linewidth]{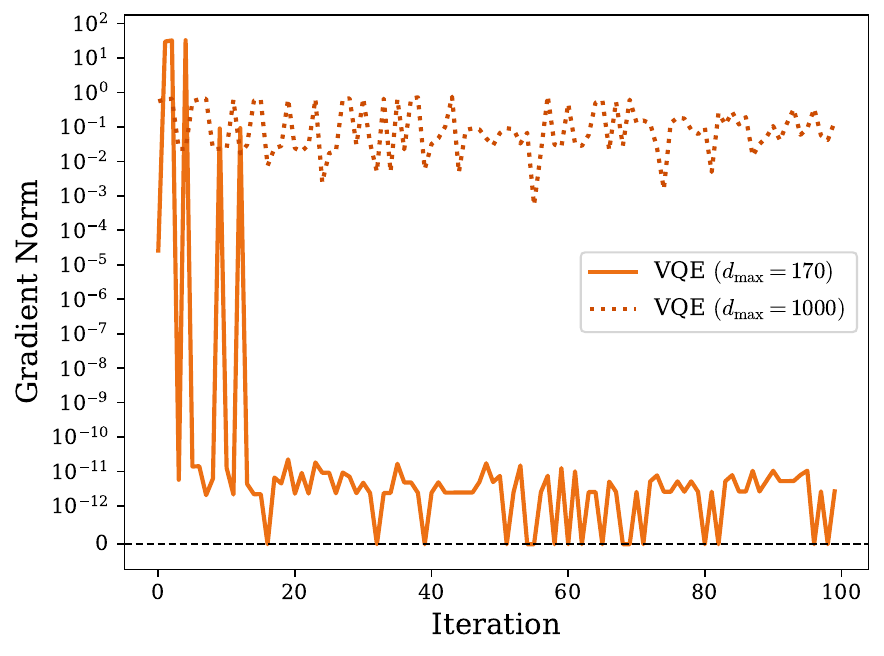}
    \end{subfigure}
    \caption{\textbf{Comparison between VQE and GQE on \boldmath$\mathrm{N}_2(10e,8o)$\unboldmath\ at \boldmath$R_{\mathrm{NN}}=2.5$~\AA{}\unboldmath.} Left: best-so-far QSCI energy error versus iteration for $d_{\max}=170$ and $1000$. Right: $\ell_2$ norm of the approximate GSPGS gradient used in the VQE optimization. Both methods are compared under the same per-iteration budget of ten QSCI evaluations. The blue band indicates chemical precision.}
    \label{fig:vqe_gqe}
\end{figure}

We use the same setting as in the optimization-history experiment of the main text, namely N$_2$ in STO-3G with the $(10e,8o)$ active space at the stretched geometry $R_{\mathrm{NN}}=2.5$~\AA{}. For GQE, each iteration samples $M=10$ circuits of length $L=10$, and their QSCI energies are used as rewards in the group relative policy optimization (GRPO) update. For the VQE baseline, we optimize a local unitary cluster Jastrow (LUCJ) circuit~\cite{Matsuzawa2020Jastrow,Motta2023LUCJ} initialized by embedding coupled-cluster singles-and-doubles (CCSD) amplitudes~\cite{Bartlett2007CCReview}. The QSCI energy obtained from this parameterized LUCJ state is minimized with a generalized simultaneous perturbation-based gradient search (GSPGS) estimator~\cite{Pachal2025GSPGS}, implemented here as the average of five symmetric simultaneous-perturbation gradients. Writing the QSCI objective as $E_{\mathrm{QSCI}}(\bm{\theta};d_{\max})$, we draw five independent Rademacher perturbation vectors $\bm{\Delta}^{(r)}_t\in\{\pm 1\}^p$ at iteration $t$ and form
\begin{equation}
\hat{\bm g}^{(r)}_t
=
\frac{E_{\mathrm{QSCI}}(\bm{\theta}_t+c_t\bm{\Delta}^{(r)}_t;d_{\max})
-
E_{\mathrm{QSCI}}(\bm{\theta}_t-c_t\bm{\Delta}^{(r)}_t;d_{\max})}{2c_t}
\,\bm{\Delta}^{(r)}_t,
\qquad r=1,\ldots,5.
\end{equation}
The gradient estimator is then averaged as
\begin{equation}
\hat{\bm g}_t=\frac{1}{5}\sum_{r=1}^{5}\hat{\bm g}^{(r)}_t,
\qquad
\bm{\theta}_{t+1}=\bm{\theta}_t-\eta_t\hat{\bm g}_t.
\end{equation}
To keep the comparison fair, one VQE iteration therefore uses $2\times 5=10$ QSCI evaluations, exactly matching the $M=10$ QSCI evaluations in one GQE iteration. We used standard simultaneous-perturbation decay schedules~\cite{Spall1992SPSA,Pachal2025GSPGS},
\begin{equation}
\eta_t=\frac{0.1}{(t+A+1)^{0.602}},
\qquad
c_t=\frac{0.05}{(t+1)^{0.101}},
\end{equation}
where the offset parameter was set to $A=10$. All other settings, including the measurement budget for each QSCI evaluation, symmetry completion, and classical diagonalization, were kept identical between VQE and GQE. We report results for $d_{\max}=170$ and $1000$ with $N_{\mathrm{iter}}=100$, and Figure~\ref{fig:vqe_gqe} shows the best-so-far energy error up to each iteration.

Figure~\ref{fig:vqe_gqe} shows that GQE yields lower final errors than VQE for both subspace budgets. The contrast is especially sharp at $d_{\max}=170$: the VQE curve improves only during the first few iterations and then remains almost unchanged around $10^{-2}$~Ha, whereas the GQE curve continues to decrease and reaches the chemical-precision window. At $d_{\max}=1000$, VQE does continue to improve, but its convergence is still substantially slower and its final best-so-far energy remains above the corresponding GQE result. These data suggest that, for QSCI ansatz optimization, searching over a discrete operator pool can be considerably easier than optimizing a continuous circuit-parameter landscape.

The gradient history in the right panel provides a direct explanation for the poor VQE performance at small $d_{\max}$. For $d_{\max}=1000$, the norm of the approximate gradient stays finite throughout the run, typically fluctuating between about $10^{-2}$ and $10^{0}$. For $d_{\max}=170$, by contrast, the gradient norm collapses to roughly $10^{-11}$ within the first $\sim 15$ iterations and repeatedly hits numerical zero thereafter. Once this happens, the best-so-far energy in the left panel also stops improving. A natural explanation is that, under a tight determinant budget, small perturbations of the LUCJ parameters often do not change the most frequently sampled determinants, so the truncated subspace selected for diagonalization remains unchanged. The QSCI objective is then effectively piecewise constant in a neighborhood of the current parameters, and the simultaneous-perturbation estimator returns an almost vanishing gradient.

This mechanism is distinct from the usual concentration-of-measure barren plateau in smooth expectation-value VQE landscapes~\cite{Cerezo2021BP}, because here the difficulty is induced by sampling and hard subspace truncation rather than by the expressibility of the ansatz alone. Nevertheless, the optimization consequence is the same: continuous-parameter updates become uninformative. Although a different ansatz family or a retuned GSPGS schedule could partly mitigate this failure mode, the present results indicate that VQE-based ansatz optimization for QSCI is substantially more fragile to a hard $d_{\max}$ constraint than GQE. In this sense, the proposed discrete operator-pool search is not only effective in the main-text comparisons against LUCJ and time-evolved baselines, but also more robust than direct VQE optimization under the same QSCI evaluation cost.

\subsection{Rotation-gate efficiency}
\label{app:rotation_gate_efficiency}

The main text uses the two-qubit-gate count as the primary cost metric because it is the most relevant proxy for near-term circuit depth and error accumulation. In an early fault-tolerant quantum computing (FTQC) setting, however, the more relevant metric is often the non-Clifford cost of implementing the state-preparation circuit. Arbitrary-angle single-qubit rotations are not native Clifford operations; they must be synthesized over a Clifford+$T$ gate set or implemented using magic-state-based protocols, and the corresponding non-Clifford resources are a central overhead in FTQC~\cite{Campbell2017Roads,RossSelinger2016}. For this reason, fault-tolerant resource studies in quantum chemistry commonly track $T$-gate or non-Clifford counts as primary cost metrics~\cite{Casares2022TFermion,Kim2022FaultTolerant}. In the present work we do not fix a particular synthesis accuracy or compilation algorithm for each rotation. Instead, we compare the number of rotation gates before synthesis, which provides an architecture-agnostic proxy for the eventual fault-tolerant non-Clifford cost. For a fixed synthesis accuracy per gate, the $T$ count of an arbitrary rotation scales only logarithmically with the target precision~\cite{RossSelinger2016}, so the pre-synthesis rotation count still captures the dominant ranking between ansatz families.

\begin{figure}[htbp]
    \centering
    \includegraphics[width=\linewidth,page=1]{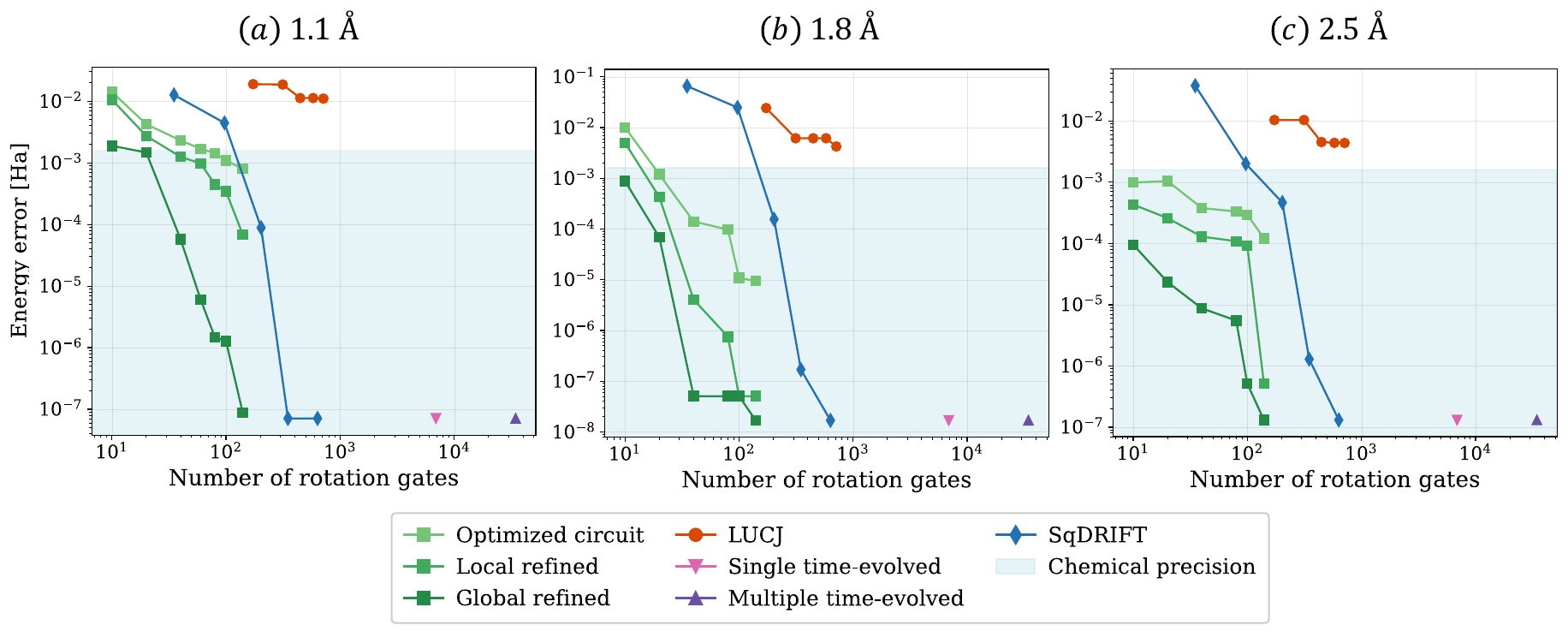}
    \caption{\textbf{Rotation-gate efficiency for \boldmath$\mathrm{N}_2 (10e,8o)$\unboldmath.} Energy error is plotted against the number of single-qubit rotation gates for $R_{\mathrm{NN}}=(a)\ 1.1$~\AA{}, $(b)\ 1.8$~\AA{}, and $(c)\ 2.5$~\AA{}. The compared state-preparation families, optimization scan, and total shot budgets are the same as in Fig.~\ref{fig:gate_efficiency}. Because refinement is entirely classical, the local- and global-refinement curves inherit the rotation-gate count of the underlying optimized circuit.}
    \label{fig:rotation_gate_efficiency}
\end{figure}

Figure~\ref{fig:rotation_gate_efficiency} shows that the gate-efficiency advantage of the proposed workflow is preserved, and becomes even clearer, under this fault-tolerant proxy. Across all three bond lengths, chemical precision is reached with only a few tens to $10^2$ rotation gates for the optimized and refined circuits. LUCJ still fails to reach chemical precision within the scanned range, while SqDRIFT typically requires several hundred rotation gates. The single- and multiple-time-evolved baselines lie much farther to the right, indicating that their good sampling performance is obtained only at a substantially larger non-Clifford implementation cost. Thus, the advantage of the learned ansatz is not specific to the NISQ-oriented two-qubit-gate metric: it persists when the same circuits are viewed through a rotation-gate proxy more directly connected to early-FTQC resource costs.

\end{document}